\documentclass[aps,twocolumn,nofootinbib,showpacs,final,balancelastpage,superscriptaddress]{revtex4}
\usepackage{graphicx}
\usepackage{tabularx}
\usepackage{amssymb}
\usepackage{amsmath}
\usepackage{amsbsy}
\usepackage{comment}
\usepackage{mathrsfs}
\usepackage{dsfont}
\usepackage{nicefrac}
\usepackage{wrapfig}
\usepackage{amsfonts}
\usepackage{url} 
\usepackage[nolist]{acronym}
\usepackage{nicefrac}
\usepackage{subfigure}
\usepackage{epsf,color,colordvi,pifont}
\usepackage{showkeys}
\DeclareFontFamily{OT1}{pzc}{}
\DeclareFontShape{OT1}{pzc}{m}{it}{<-> s * [1.10] pzcmi7t}{}
\DeclareMathAlphabet{\mathpzc}{OT1}{pzc}{m}{it}
\def\dbar{{\mathchar'26\mkern-12mu \textrm{d}}}

\newcommand{\vecAA}{\pmb{\mathpzc A}}

\newcommand{\Nplat}{N_{\mathrm{plateau}}}

\begin{document}

\begin{acronym}[ABCD]
\acro{PP}{pair production}
\acro{bOEF}{bifrequent oscillating electric field}
\acro{ODEs}{Ordinary Differential Equations}
\end{acronym}

\title{Low-dimensional approach to pair production in an oscillating electric field:  Application to bandgap graphene layers}

\author{I. \surname{Akal}}
\altaffiliation{New address: Theory Group, Deutsches Elektronen-Synchrotron DESY, Notkestr. 85, 22603 Hamburg, Germany}
\author{R. \surname{Egger}}
\author{C. \surname{M\"{u}ller}}
\author{S. \surname{Villalba-Ch\'avez}}
\email{selym@tp1.uni-duesseldorf.de}
\affiliation{Institut f\"{u}r Theoretische Physik, Heinrich-Heine-Universit\"{a}t D\"{u}sseldorf, Universit\"{a}tsstr.\,1, 40225 D\"{u}sseldorf, Germany}

\begin{abstract}
The production of particle-antiparticle pairs from the quantum field theoretic ground state in the presence of an external electric field is studied. 
Starting with the quantum kinetic Boltzmann-Vlasov equation in four-dimensional spacetime, we obtain the corresponding equations in lower dimensionalities
by way of spatial compactification. Our outcomes in $2+1$-dimensions are applied to bandgap graphene layers, where the charge carriers have the particular
property of behaving like light massive Dirac fermions. We calculate the single-particle distribution function for the case of an electric field oscillating in 
time and show that the creation of particle-hole pairs in this condensed matter system closely resembles electron-positron pair production by the Schwinger effect.
\end{abstract}



\date{\today}

\maketitle

\section{Introduction}

Attaining the vacuum instability in a strong, macroscopically extended electromagnetic field through spontaneous creation of electron-positron pairs represents a major incentive for quantum electrodynamics (QED) 
and various branches of high-energy physics. For the case of a static electric field, the corresponding Schwinger rate of pair production (PP) has the form $\dot{\mathpzc{N}}\sim \exp(-\pi E_{c}/E)$ \cite{Sauter:1931,Heisenberg:1935,Schwinger:1951nm}, 
where $E$ is the applied field and $E_{c}=m^2c^3/(\vert e\vert\hbar)$ the critical field of QED.\footnote{Here and henceforth $m$ and $\vert e\vert$ are the electron mass and its absolute charge, respectively. 
The  speed of light in vacuum and the Planck constant  will be  denoted by $c$ and $\hbar$, respectively.} Due to its non-analytic field dependence, the Schwinger effect exhibits a manifestly nonperturbative character. 
Its experimental observation, while being highly desirable, has been prevented until now by the huge value of $E_{c}\sim 10^{16}\ \rm V/cm$ which is not accessible in the laboratory yet. It would not only 
verify a central prediction of QED, but also have implications for various other phenomena which share its  principal features \cite{Casher:1978wy,rafelski:1978,Kachelriess,Kharzeev}.

Substantial efforts are being spent to bring the Schwinger effect into experimental reach. The next generation of high-intensity laser facilities currently under construction is expected to reach peak field intensities 
on the order of $\sim10^{25}\,\rm W/cm^2$ \cite{ELI, xcels}. The corresponding electric fields, having micrometer extensions and femtosecond durations, reach the percent level of the critical field $E_{c}$. To 
further increase the chance for an experimental realization of Schwinger PP, theoreticians have proposed to superimpose multiple laser beams \cite{Bulanov2010}, exploit the Lorentz boost inherent to 
relativistic particle-laser beam collisions \cite{Burke:1997ew,Muller:2003,Sieczka2006} or enhance the rate by the aid of an assisting high-frequency (photon) field 
\cite{dgs2009,DiPiazza:2009py,Orthaber2011,Grobe2012,Jansen2013,Akal:2014eua}. 

While these sorts of facilitated Schwinger PP are very promising, they do not constitute the only way of probing the vacuum instability. Alternatively, one may follow a complementary route by inspecting physical 
environments where analogues of the Schwinger effect arise whose detection might be easier. Since its recent  discovery, graphene \cite{GrExp1,GrExp2,GrExp3,Neto,kotov,basov} has become an ideal candidate for such a purpose. 
Fermionic quasiparticles close to one of the two $K_\pm$ points in this  two dimensional monolayer of carbon atoms  behave like relativistic particles \cite{Wallace}. They feature a dispersion relation like Dirac fermions, 
with the speed of light $c$ being replaced by the Fermi velocity $\mathpzc{v}_f\approx c/300$,  and the eigenstates have a Dirac spinor structure in sublattice space. As a consequence, fundamental phenomena of relativistic 
quantum physics find their low-energy counterpart in graphene, such as Klein tunneling  \cite{Klein}, Coulomb supercriticality \cite{kotov}, many-body renormalization effects \cite{basov}, or universal scaling 
phenomena \cite{KloepferSuper}.

Also the analogue of Schwinger PP, i.e., field-induced creation of particle-hole pairs via low-energy electronic excitations from valence to conduction band in graphene has been studied theoretically. The process was 
examined by methods of $2+1$-dimensional QED in a constant electric background \cite{Allor:2007ei},\footnote{In this background  field, various  related effects  which might  provide experimental signatures of the 
production of electron-hole pairs  have been investigated \cite{Lewkowiczprl,Lewkowiczprb,yokomizo,dora1,dora2}. } a  Sauter-like field  \cite{Mostepanenko}, and an  electric field oscillating in time \cite{Avetissian,Fillion-Gourdeau:2015dga}.  
The lack of an energy gap at the $K_\pm$ points in graphene implies, however, that the charge carriers behave like massless Dirac fermions. This property prevents the existence of a critical field in graphene and, thus, the exponential 
suppression of the production rate which represents a main characteristic of the original Schwinger effect. 

There are various techniques to induce a bandgap $\Delta\varepsilon$ in graphene, e.g., by epitaxial growth on suitable substrates \cite{Substrate1,Substrate2}, elastic strain \cite{basov}, or Rashba spin splittings on 
magnetic substrates \cite{Varykhalov}.  
Then a non-zero mass $\mathpzc{m}=\Delta\varepsilon/(2\mathpzc{v}_f^2)$  is associated with the charge carriers whose high mobility still resembles the one of relativistic particles as long as their momenta lie below 
$\sim 3 \ \mathrm{eV}/\mathpzc{v}_f$ \cite{Neto}. In this situation, the phenomenology of  the field-induced particle-hole production process can be expected to show characteristics much closer to the electron-positron PP in QED. 
Such an analogy  offers an ideal opportunity of getting valuable insights about the original Schwinger mechanism in an experimentally accessible low-energy domain. This is the motivation for the present study.


In this paper, we calculate the production of pairs of massive Dirac particles from the quantum field theoretic ground state in an external electric field. Starting from the well-known QED description of the process in 
$3+1$ dimensions, we develop  a formalism in reduced dimensionality which is applicable to bandgap graphene layers. Analytical results for the production rate in a Constant Electric Field (CEF) are obtained. For an electric field 
oscillating in time, the production rate and the momentum distribution of created particles are examined by numerical calculations. We demonstrate that--similar to PP  in the unstable QED vacuum--the 
production of  electron-hole pairs in a vicinity of the Dirac points in graphene layers  is governed by a kinetic equation  characterized  by a non-Markovian nature and describe suitable field parameters for 
experimental observation of the effect.

Our paper is organized as follows. In  Sec.~\ref{sec:QKA} we describe  some basic features of the PP process in an external electric field in $3+1$-dimensional QED. Afterwards, by compactifying spatial dimensions, the quantum 
kinetic Boltzmann-Vlasov equations  in $2+1$ and $1+1$ dimensions are derived. The outcomes of this analyses are first applied to the case of a constant external electric field, where agreement with previous studies is found. 
This part is followed by a subsection which exposes the details necessary for applying  the kinetic equation in $2+1$-dimensions to graphene layers.  The  numerical results are presented in Sec.~\ref{sec:numerical} for the 
particular situation in which the  electric field oscillates in time. We discuss the resonant effects and Rabi-like  oscillations  which arise in the  single-particle distribution function. The density of pairs yielded is also  
investigated. Further comments and remarks are  given in the conclusion.

\section{Quantum Kinetic Approach} \label{sec:QKA}

\subsection{General remarks in 3+1-dimensions \label{subsec:boltzman-vlasov}}

First,  we  shall  investigate the production of pairs in a field as can be conceived from laser waves. However, its description represents  a major task  from both  analytic and numeric viewpoints.   
Difficulties, introduced by the unavoidably complicated nature of such a  wave,  are frequently  simplified  for the sake  of having a  concise analytical description of the production 
process.\footnote{Such simplifications cannot reduce the external field  to a plane-wave, because in this case  the field invariants $\mathscr{F}=(E^2-B^2)/2$ and $\mathscr{G}=\pmb{E}\cdot\pmb{B}$ 
vanish identically, and the vacuum-vacuum transition amplitude does not decay against  electron-positron pairs \cite{Fradkin,Dunne:2004nc}.} A first step towards this direction  
is  taken  by considering  an electric-like background  which can be obtained--to a good approximation--through  a head-on  collision of  two linearly polarized  
laser pulses  with  equal intensities, frequencies  and  polarization directions. The  field which results from this procedure  is  a spatially  inhomogeneous standing  electromagnetic wave  depending  
upon  the temporal coordinate. However,  for the  numeric treatment, the  dependence on the space coordinates  still constitutes an important issue which is frequently  avoided, although 
there exist already some first results  including their effects \cite{ruf2009zz,Hebenstreit:2010vz}. Because of this, most of the  studies  developed in this research area  consider  the simplest  model  in 
which the background is a  homogeneously distributed electric field oscillating in time \cite{Grib:1972te,popov,Nikishov1,Brezin,Bagrov,Gitman,Salamin}. 

There exist various theoretical approaches for describing the spontaneous  production  of electron-positron pairs in an external electromagnetic field. Many of them rely on the theoretical 
framework  of QED in unstable vacuum \cite{Fradkin}. In the case of a spatially  homogeneous but time dependent electric field $\pmb{E}(t)=(0,E(t),0)$,\footnote{Hereafter we  assume  that  the external 
field  is not affected by the process occurring in it. Therefore, we fully disregard the potential  realization of avalanche  processes \cite{Fedotov:2010ja}.} such a  framework   allows us to undertake  
the problem  either  through  the  S-matrix formalism \cite{Mostepanenko,Nikishov1,popov,Bagrov,Brezin,Gitman,Narozhny,Gavrilov:1996pz} or from  the transport theory point of view 
\cite{Schmidt:1998vi,Schmidt:1999vi,alkofer:2001ib,Bloch:1999eu,vinnik,Hebenstreit:2009km,Kohlfurst:2012rb,Akal:2014eua,Vinnik:2001qd,BlaschkeCPP}. Both formulations are equivalent and complement each other. 
In this paper,  we adopt the latter   description  
which  underlines  the  nonequilibrium nature of the PP.  In this context,  its  investigation is  carried out in terms of  the  distribution function  $W(\pmb{p},t)$ of electrons and  
positrons  to which the degrees of freedom in the external field  are relaxed  at asymptotically large times  [$t\to\pm\infty$],  i.e., when  the electric field is switched off $\pmb{E}(\pm\infty)\to 0$. 
The time evolution  of this quantity is dictated  by  a quantum  kinetic equation involving  a pronounced non-Markovian feature:\footnote{In the 
following we set the Planck constant   equal to unity, $\hbar=1$.}
\begin{eqnarray}\label{vlasov}
&&\dot{W}(\pmb{p},t)=\frac{e E(t) \epsilon_\perp c}{\mathpzc{w}_{\pmb{p}}^2(t)}\int_{-\infty}^t dt^\prime \frac{e E(t^\prime)\epsilon_\perp c}{\mathpzc{w}_{\pmb{p}}^2(t^\prime)}
[1-W(\pmb{p},t^\prime)]\nonumber\\&&\qquad\quad\qquad\times\cos\left[2\int_{t^\prime}^t dt^{\prime\prime}\ \mathpzc{w}_{\pmb{p}}(t^{\prime\prime})\right],
\end{eqnarray}and in which the  vacuum initial condition $W(\pmb{p},-\infty)=0$ is  assumed.  This equation represents  a semiclassical approximation in  the sense that the external background 
field is not quantized while the equation itself  results from the quantization of the Dirac field. Note that, hereafter, a dot  indicates a total time derivative. We also point out  that  $W(\pmb{p},t)$  
involves  a sum over both spin states,  providing  an overall  factor two. The above  formula is characterized by  the transverse
energy squared  $\epsilon_\perp^2=m^2c^4+\pmb{p}_\perp^{\,2}c^2$ and  the total energy squared   $\mathpzc{w}^2_{\pmb{p}}(t)=\epsilon_\perp^2+[p_\parallel-e\mathpzc{A}(t)/c]^2c^2$, 
with  $\pmb{p}_\perp=(p_x,0,p_z)$ and $\pmb{p}_\parallel=(0,p_y,0)$ being  the components of the canonical momentum  perpendicular and parallel 
to the direction of the field, respectively. Note that the  four-potential  of the external field  $\mathpzc{A}_\mu(t)$  is chosen in  the temporal 
gauge [$\mathpzc{A}_0(t)=0$]  so that $E(t)=-(1/c) d\mathpzc{A}/dt$.   
 
Further comments are in order. Firstly, the quantum Vlasov equation [Eq.~(\ref{vlasov})]  does  not take into account neither the collision between 
the created particles nor  their inherent  radiation  fields. In the presence of a CEF both phenomena are predicted to become  relevant  
as  the field strength $E$ reaches the critical scale of QED $E_c= m^2c^3/\vert e\vert$  \cite{Tanji:2008ku,Bloch:1999eu,Vinnik:2001qd}.  
So,  the solution of Eq.~(\ref{vlasov}) is expected to be  trustworthy in the 
subcritical regime $E\ll E_c$ where the number of  produced pairs per unit of volume reads 
\begin{equation}
\mathpzc{N}_{\ 3+1}=\lim_{t\to\infty}\int \dbar^3 p \, W(\pmb{p},t),
\label{eqn:tot-no}
\end{equation}with  the  shorthand notation $\dbar\equiv d/(2\pi)$.

Finally, we remark that,  when the external field is oscillating in time with frequency $\omega$ and  amplitude  $E$,  the resulting $\mathpzc{N}_{\ 3+1}$  should  be  much  smaller 
than the maximum  density $\mathpzc{N}_{\ \mathrm{max}}\sim E^2/(2\omega)$  that can be created  from it, otherwise the external field model is no longer justified.

\subsection{Boltzmann-Vlasov equations in low dimensional spacetimes \label{subsec:vlasov-lowdimensionality}}

Quantum kinetic theory constitutes an appropriate approach for investigating the production of electron-positron pairs by a time-dependent electric field in low dimensional spacetimes.  
The corresponding  quantum Vlasov equations can be  derived  from the respective  Dirac equation in the field by  applying a method similar to the  
one used  in the determination  of  Eq.~(\ref{vlasov}) [for  details,  we refer the reader to  Refs.~\cite{Schmidt:1998vi,Schmidt:1999vi}].  
Alternatively, a  dimensional  reduction   $\grave{a}\ la$ Kaluza-Klein \cite{kaluza,klein}   can be  carried out on the   $3+1$  quantities, particularly, on the  coordinates perpendicular 
to the field.\footnote{This requirement is in connection with the $2+1$ dimensional version of QED in which  the electric field lies in the plane defined by 
the space-coordinates.} By using the  latter procedure,  we  obtain  formalisms of different dimensionality. 

The first step allows us to describe  the  production of pairs in a Minkowski space with $2+1$-dimensions $\rm M_{2+1}$. In order to show this,  we take  the dimension in excess $x^i$ to be curled up into a  circle  
$\rm S^{1}$ with a radius $\mathpzc{R}$. In this context the motion of  the Dirac fermions is confined  to the  interval $0\leqslant x^i\leqslant2\pi\mathpzc{R}$, 
and since  the compactified coordinate  is  periodic and the electric field is homogeneous, one can  Fourier expand the Dirac field in terms of the  corresponding  
quantized momentum  $ p_{\mathpzc{n}}^i=\mathpzc{n}\mathpzc{R}^{-1}$ with $\mathpzc{n}\in \mathbb{Z}$. The exponentials involved  in this expansion $\sim\exp(i\mathpzc{n} x^i/\mathpzc{R})$ 
heavily  oscillate as the limit $\mathpzc{R}\to 0$ is taken into account. In such  circumstances,  only  the fundamental mode  $\mathpzc{n}=0$--corresponding 
to vanishing momentum--dominates. Hence,  once  the  momentum $\pmb{p}$ is  locked up to a plane perpendicular to $\pmb{x}^i$, the spontaneous production of 
electron-positron pairs driven by an Oscillating Electric Field (OEF) in a  $2+1-$dimensional spacetime is effectively described by  Eq.~(\ref{vlasov}). However, 
unlike the case  in  $3+1$-dimensions, the single-particle distribution function  in a $2+1$-dimensional spacetime  does not involve a summation over the double-valued  
spin indices. We remark that the contribution of this sum is canceled out by dividing the resulting expression for  $W(\pmb{p},t)$ by a factor $2$ \cite{Hebenstreit:2010vz,Akal:2014eua}. 

It is worth mentioning that the results which follow from this effective description will be valid as long as the typical energy involved in the problem is much below the
characteristic scale $\varepsilon_0\simeq  c \mathpzc{R}^{-1}$. On the other hand, the expression for the number of produced pairs per unit of area in $\rm M_{2+1}$  results from Eq.~(\ref{eqn:tot-no}) 
by performing a  transition to the discrete  limit $\int \dbar p\to \frac{1}{2\pi\mathpzc{R}}\sum_{p^i}$ in which only the contribution 
of the vanishing  mode must be considered.  Accordingly,  
\begin{equation}\label{numberparticles21}
\mathpzc{N}_{\ 2+1}\equiv\frac{1}{2}\lim_{\mathpzc{R}\to0} 2\pi \mathpzc{R}\ \mathpzc{N}_{\ 3+1}= \lim_{t\to\infty}\int \dbar^2 p \, W_{2+1}(\pmb{p},t),
\end{equation}where $W_{2+1}(\pmb{p},t)$ refers to the single-particle distribution function $W(\pmb{p},t)$ divided by $2$ and  with  the momentum component  $p^i$ having  been set to zero. 

Straightforward analyses, similar to  those  made in the previous case allow us  to apply Eq.~(\ref{vlasov}) when  the  production of pairs takes place in $1+1$-dimensions.  
In such a case, two spatial coordinates have to  be compactified on a torus $\rm S^1\times S^1$,  which requires to  set the respective components of the quantized  momentum 
to zero as the  characteristic radii of this manifold  $\tilde{\mathpzc{R}}$ and  $\mathpzc{R}$ vanish  identically. Then the  asymptotic expression $W_{1+1}(\pmb{p},\infty)$,  
resulting from the quantum Vlasov equation--after dividing by $2$--defines  the linear density of created particle pairs $\mathpzc{N}_{\ 1+1}=\int \dbar p \ W_{1+1}(\pmb{p},\infty)$.

We apply the procedure described above  by supposing  the occurrence  of the  process in $\rm M_{2+1}$  through  a constant  electric field [$\mathpzc{A}(t)=-cEt$]. Seeking for  simplicity,  
we will take advantage of the known asymptotic expression for the single-particle  distribution function in a four-dimensional spacetime $W(\pmb{p},\infty)\simeq2\exp[-\pi\epsilon_\perp^2/(\vert e\vert Ec)]$ \cite{Hebenstreit:2010vz}. 
After  compactifying  the axis lying  perpendicular to the external field and  considering the  fundamental contribution [$p_z=0$],  we find  that the rate of PP  per unit area   is
\begin{equation}
\dot{\mathpzc{N}}_{\ 2+1}\approx\frac{(\vert e\vert E)^{\nicefrac{3}{2}}}{4\pi^2c^{\nicefrac{1}{2}}}\exp\left(-\pi\frac{E_c}{E}\right),\label{ratenumber21}
\end{equation}provided $E\ll E_c$. A similar outcome results  in  $1+1$-dimensions. In this case we obtain that the 
asymptotic expression of the single-particle distribution function is momentum independent $W_{1+1}(\infty)\simeq\exp[-\pi E_c/E]$ and the rate of yielded particle pairs 
per unit of length reads
\begin{equation}
\dot{\mathpzc{N}}_{\ 1+1}\approx\frac{\vert e\vert E}{2\pi}\exp\left(-\pi\frac{E_c}{E}\right).\label{ratenumber11}
\end{equation}
The rates presented in Eq.~(\ref{ratenumber21}) and (\ref{ratenumber11}) coincide with those previously obtained in Refs.~\cite{Gavrilov:1996pz,Kim:2000un,Cohen:2008wz,Hebenstreit:2013qxa} by   other  methods.  
A comparison between both expressions  reveals a clear dependence on the respective spacetime topology.

\subsection{Extension to  Dirac fermions in  graphene layers \label{subsec:res-effects}}

We wish to apply the procedure described so far to the production of Dirac fermions in a graphene layer. Since this kind of layer approaches to  a  
$2+1$-dimensional system, one could expect that the results  previously derived  are  suited for such a purpose. However, inherent features of this  
material prevent us to  proceed in a straightforward way. While some of these characteristics can be incorporated, there are other details which 
require certain attention.  For instance,  our previous expressions do not take into account effects resulting  from  finite temperatures. 
Accordingly, it must be  understood that their applicability will be trustworthy in the zero temperature limit. Also, we will suppose that the  
electron-hole symmetry is preserved in this layer,  a fact theoretically  verified within the nearest-neighbours tight-binding model but no longer 
true as the  next-to-nearest-neighbours interactions are considered  \cite{Wallace}. 

In contrast to previous investigations \cite{Allor:2007ei,Mostepanenko,Avetissian,Fillion-Gourdeau:2015dga}, here and in the following the charge carriers will be considered  
with a tiny  mass $\mathpzc{m}$ corresponding to the gap $\Delta\varepsilon=2\mathpzc{m}\mathpzc{v}_f^2$. In the numerical calculations in Sec.~\ref{sec:numerical}  
the specific value  $\Delta\varepsilon=0.26\ \rm eV$  will be chosen for practical purposes;  such an energy gap can originate, e.g., from   the epitaxial  growth 
of graphene on SiC substrates \cite{Substrate1}. For comparison the energy gap $\Delta\varepsilon=0.12\ \rm eV$ will be considered  in addition.
We note that also other values of the energy gap can be induced in graphene \cite{Substrate2}.

Observe that, when adapting the  $2+1$-dimensional version of Eq.~(\ref{vlasov}),  we have to take into account  that--in graphene--the Fermi velocity $\mathpzc{v}_f\approx c/300$ cannot be exceeded. The inclusion 
of this constraint demands that one slightly  modifies the quantum Boltzmann-Vlasov equations in $\rm M_{2+1}$, since it characterizes the instability of the  
vacuum where the created electrons and positrons  might travel with velocities greater than $\mathpzc{v}_f$ but always smaller than $c$. Such modifications 
are carried out by replacing $mc^2\to \mathpzc{m}\mathpzc{v}_f^2$, $p_i c\to p_i \mathpzc{v}_f$, $eEc\to eE\mathpzc{v}_f$, $\mathpzc{A}c\to \mathpzc{A}\mathpzc{v}_f$  
so that the production of quasiparticle-hole pairs in graphene turns out to be  governed by 
\begin{eqnarray}\label{vlasovgraphene}
&&\dot{W}_{\mathpzc{g}}(\pmb{p},t)=Q(\pmb{p},t)\int_{-\infty}^t dt^\prime Q(\pmb{p},t^\prime)
\left[\frac{1}{2}-W_{\mathpzc{g}}(\pmb{p},t^\prime)\right]\nonumber\\&&\qquad\quad\qquad\times\cos\left[2\int_{t^\prime}^t dt^{\prime\prime}\ \mathpzc{w}_{\pmb{p}}(t^{\prime\prime})\right],
\end{eqnarray}where the function $Q(\pmb{p},t)\equiv e E(t)\mathpzc{v}_f \epsilon_\perp /\mathpzc{w}_{\pmb{p}}^2(t)$ has been introduced. While  $\epsilon_\perp^2=\mathpzc{m}^2\mathpzc{v}_f^4+p_x^{\,2}\mathpzc{v}_f^2$ 
is the transverse energy squared of the Dirac fermions,  $\mathpzc{w}^2_{\pmb{p}}(t)=\epsilon_\perp^2+[p_y-e\mathpzc{A}(t)/c]^2\mathpzc{v}_f^2$ is their respective  total 
energy squared.  

Further comments are in  order. Firstly,  Eq.~(\ref{vlasovgraphene}) shows  that the production of electron-hole pairs in graphene is a nonequilibrium phenomenon. Observe that--as in QED--the 
combination of the nonlocality  in time and the memory effects closely associated with the quantum  statistic factor $\sim \left[1/2-W_\mathpzc{g}(\pmb{p},t)\right]$ provides  Eq.~(\ref{vlasovgraphene})  
with a pronounced  non-Markovian feature \cite{Schmidt:1998vi,Bloch:1999eu,vinnik}.  It means that the single-particle distribution function $W_\mathpzc{g}(\pmb{p},t)$ depends on 
the number of electron-hole pairs  already present in the system. We remark that the spectral information encoded in  Eq.~(\ref{vlasovgraphene}) will be  valid in a vicinity of any 
of the two inequivalent points in the reciprocal space $\pmb{K}_\pm$. Accordingly,  the momentum of the quasiparticles $\pmb{p}$  must be understood as  relative to $\pmb{K}_\pm$  
with  $\vert\pmb {p}\vert\ \ll \vert\pmb{K}_\pm\vert=\frac{4\pi}{3\sqrt{3}a_0}$ and  $a_0=0.142\ \rm nm$ \cite{Neto}.  It is, indeed,  the previous restriction that does guarantee the relativistic-like 
behavior of  the charge carriers. Thus, in calculating the  density of pairs  per unit of area [Eq.~(\ref{ratenumber21})]  the respective integral must  be carried out  over a surface  limited 
by $p_{\mathrm{max}}\ll \vert\pmb{K}_\pm\vert\sim 3\ \mathrm{eV}/\mathpzc{v}_f$. As such, the existence of the two inequivalent Dirac points,  in combination with the spin degeneracy,  leads to four 
kinds of quasiparticles. Therefore, the total number density of  produced particle pairs in graphene is
\begin{equation}
\mathpzc{N}_{\mathpzc{g}}=4\int_{\vert\pmb {p}\vert\ \ll \vert\pmb{K}_\pm\vert} \dbar^2p\  W_{\mathpzc{g}}(\pmb{p},\infty).\label{Ngraphene}
\end{equation}

The application of Eq.~(\ref{Ngraphene}) to the case driven by a CEF leads to an expression for $\mathpzc{N}_{\mathpzc{g}}$ that  differs from the one which would result  
from Eq.~(\ref{ratenumber21}):\footnote{The reader interested in the details of this integration may find it  helpful  to refer  to Sec.~V of Ref.~\cite{Mostepanenko}, where it is explained  
in detail.} 
\begin{eqnarray}\label{pairsingraphene}
\mathpzc{N}_{\mathpzc{g}}\approx\frac{2p_{\mathrm{max}}}{\pi^2}\left(\frac{\vert e\vert E}{\mathpzc{v}_f}\right)^{\nicefrac{1}{2}}\exp\left(-\pi\frac{E_\mathpzc{g}}{E}\right),
\end{eqnarray}where  $E_\mathpzc{g}=\mathpzc{m}^2\mathpzc{v}_f^3/\vert e\vert$ is the critical field in graphene. For example, assuming an energy gap of  $\Delta\varepsilon=0.26\ \rm eV$, 
its value is $E_\mathpzc{g}\simeq 2.6 \times 10^5\ \rm V/cm$.  It may be seen as arising from  the break down of the chiral symmetry through the mass $\mathpzc{m}$. We note  that this critical 
scale turns out to be much smaller than the critical scale of QED  $E_c=1.3 \times 10^{16}\ \rm V/cm$ by eleven  orders of magnitude. The expression in Eq.~(\ref{pairsingraphene}) can be understood 
as a ``saturation'' density of Dirac-like particle pairs in graphene. It is reached when the interaction time with the external CEF approaches $T_{\rm sat}\sim p_{\rm max}/(eE)$. For larger interaction 
times, also particles with momenta exceeding $p_{\rm max}$ are created which are not properly described by the Dirac equation \cite{Mostepanenko,Fillion-Gourdeau:2015dga}.

Finally, it must be understood that the effective Boltzmann-Vlasov equations describing the spontaneous creation  
of electron-hole pairs apply as long as the  electric  field  satisfies the condition $E\ll E_\mathpzc{g}$.  As in the vacuum case, Eq.~(\ref{vlasovgraphene}) does not take into account neither the 
effects coming from the inherent radiation of the charge carriers nor the collisions between the created quasiparticles. 


\section{Numerical  results in an electric field oscillating in time}
\label{sec:numerical}

\subsection{Resonant approach and numerical aspects}

The similarity between the kinetic equation describing the PP in graphene [Eq.~(\ref{vlasovgraphene})] and the one corresponding to  $3+1$ dimensional QED [Eq.~(\ref{vlasov})]  
allows us to extrapolate  various  outcomes associated with the production of electron-positron pairs to the graphene scenario. For instance, if the electric 
field oscillates  in time periodically,  we expect that  $W_{\mathpzc{g}}(\pmb{p},t)$ resembles the characteristic resonances  associated with the absorption of  multiple 
energy packages [``photons'']  from the field \cite{popov,Mostepanenko:1974im,Narozhny,Avetissian2,mocken2010,BlaschkeCPP}.  This  phenomenon  takes place as the resonance condition  
\begin{equation}\label{rensonantcondition}
2\bar{\varepsilon}_{\pmb{p}}\simeq n\omega
\end{equation} holds. In this relation,  $n$ denotes  the number of absorbed  photons  whereas 
$\bar{\varepsilon}_{\pmb{p}}=\frac{1}{\tau} \int_{0}^{\tau}dt\mathpzc{w}_{\pmb{p}}(t)$
refers  to the quasi-energy of the produced particles, i.e. the energy averaged over the total pulse length  $\tau$. The behavior of the distribution function $W_{3+1}(\pmb{p},t)$ 
near a resonance  characterized by $n$ is also known. Quoting the result derived in \cite{Akal:2014eua,Mostepanenko:1974im,Narozhny} and applying the procedure outlined in the previous section one has 
that
\begin{eqnarray}
W_{\mathpzc{g},n}(\pmb{p},t)\approx\frac{1}{4}\frac{\vert\Lambda_{n}(\pmb{p})\vert^2}{\Omega_{\mathrm{Rabi}}^2(\pmb{p})}\sin^2\left[\Omega_{\mathrm{Rabi}}(\pmb{p})(t-t_\mathrm{in})\right].\label{resonantdistributionfunction}
\end{eqnarray} The formula above was obtained  by supposing  that the field is suddenly turned on at $t_\mathrm{in}$ and instantaneously turned off after the interaction time. 
Here $\Lambda_{n}(\pmb{p})$  is a complex time-independent coefficient whose explicit expression  is not important here. In Eq.~(\ref{resonantdistributionfunction}),
$\Omega_{\mathrm{Rabi}}(\pmb{p})=\frac{1}{2}\left[\vert\Lambda_{n}(\pmb{p})\vert^2+\Delta_{n}^2(\pmb{p})\right]^{\nicefrac{1}{2}}$ denotes 
the  Rabi-like frequency of the vacuum with    $\Delta_{n}(\pmb{p})\equiv2\bar{\varepsilon}_{\pmb{p}}-n\omega$ being  the detuning parameter. At this point  we should mention 
that the above resonant approximation  is valid if the Rabilike frequency  is slow in comparison with the laser frequency, $\Omega_{\mathrm{Rabi}}(\pmb{p})\ll\omega$ \cite{Avetissian2,mocken2010}.

Let us now investigate the outcome resulting from a numerical evaluation  of  Eq.~(\ref{vlasovgraphene}) when the OEF is described  by  a potential of the form
\begin{align}
 \vecAA (t) = &- \frac{cE}{\omega}\mathscr{F}(\phi)\sin(\phi) \hat{\pmb{n}},
 \label{eqn:gauge-field}
\end{align}where $\omega$ and $E$  are  the frequency and the electric field amplitude, respectively. Besides, here  $\phi=\omega t$ and $\hat{\pmb{n}}^T = (0,1,0)$ defines the polarization 
direction of the field. In Eq.~\eqref{eqn:gauge-field} the envelope function is chosen with $\sin^2$-shaped turn-on and turn-off segments and a plateau region of constant field intensity in between. Explicitly,
\begin{equation}
\mathscr{F}(\phi) = \begin{cases}\sin^2\left(\frac{1}{2}\phi\right) & 0 \leqslant \phi <  \pi\\ 
1 & \pi  \leqslant \phi\leqslant 2 \pi \mathpzc{K}\\ 
\sin^2\left(N\pi - \frac{1}{2}\phi\right) & 2 \pi \mathpzc{K} < \phi \leqslant 2 \pi N\\ 
0 & \mbox{otherwise} \end{cases} ,
\label{eqn:envelope}
\end{equation}
where $N =\Nplat + 1$ and $\mathpzc{K} = N - \frac{1}{2}$ holds.  Observe that Eq.~\eqref{eqn:gauge-field} and  \eqref{eqn:envelope} guarantee the  starting of the OEF  with zero-amplitude 
at  $t=0$.

Although  Eq.~(\ref{vlasovgraphene}) provides various  physical insights inherent  to the   PP  process in graphene,  its  numerical solution is   easier  to determine when an equivalent 
system of ordinary differential equations is considered  instead:
\begin{eqnarray}\label{firstequa}
&&i\dot{f}(\pmb{p},t)=\mathpzc{a}_{\pmb{p}}(t)f(\pmb{p},t)+\mathpzc{b}_{\pmb{p}}(t)g(\pmb{p},t),\\ 
&&i\dot{g}(\pmb{p},t)=\mathpzc{b}^*_{\pmb{p}}(t)f(\pmb{p},t)-\mathpzc{a}_{\pmb{p}}(t)g(\pmb{p},t).\label{secondequa}
\end{eqnarray}In this context  the distribution function is given by $W_{\mathpzc{g}}(\pmb{p},t)=\vert f(\pmb{p},t)\vert^2$ and 
the initial conditions are chosen so that $f(\pmb{p},-\infty)=0$ and $g(\pmb{p},-\infty)=1$. The remaining  parameters contained 
in these formulas  are given by
\begin{eqnarray}\label{coefficient1}
\begin{array}{c}
\displaystyle\mathpzc{a}_{\pmb{p}}(t)=\mathpzc{w}_{\pmb{p}}(t)+\frac{eE(t)p_x\mathpzc{v}_f^2}{2\mathpzc{w}_{\pmb{p}}(t)[\mathpzc{w}_{\pmb{p}}(t)+\mathpzc{m}\mathpzc{v}_f^2]},\\ \\
\displaystyle\mathpzc{b}_{\pmb{p}}(t)=\frac{1}{2}\frac{eE(t)\mathpzc{v}_f\epsilon_\perp}{\mathpzc{w}_{\pmb{p}}^2(t)}\exp\left[-i\arctan\left(\frac{p_x q_\parallel\mathpzc{v}_f^2}{\epsilon_\perp^2+\mathpzc{m}\mathpzc{v}_f^2\mathpzc{w}_{\pmb{p}}(t)}\right)\right],\\
\end{array}\nonumber
\end{eqnarray}where  the longitudinal  kinetic momentum is $q_\parallel=p_y-e\mathpzc{A}(t)/c$.  The equivalence between  Eq.~(\ref{vlasovgraphene}) and  the system above [Eq.~(\ref{firstequa})-(\ref{secondequa})]
has been mathematically established  in several references (see for instance \cite{Hebenstreit:2010vz,Schmidt:1998vi,Akal:2014eua}). However, it is worth mentioning  that various representations of the  
Boltzmann-Vlasov equation can be found in the literature. Their use is  mainly motivated by an optimization of the numeric assessment.


\subsection{Results and discussions} \label{subsec:res}

As mentioned in the previous section, the mass of the Dirac fermions will be taken as $\mathpzc{m}=\Delta\varepsilon/2\mathpzc{v}_f^2$, which  corresponds  to $\mathpzc{m}\approx 5.4\ \mathrm{keV}/c^2$ for $\Delta \varepsilon=0.12\ \rm eV$ and $\mathpzc{m}\approx 11.7\ \mathrm{keV}/c^2$ for $\Delta \varepsilon=0.26\ \rm eV$.  
We shall set the field  frequency  to $\omega =24.032\ \rm meV$. The plateau region will comprise $\Nplat=241$ cycles so that the total pulse length is $\tau=2\pi N/\omega\simeq41.625\ \rm ps$. 
On the other hand, the field strength is taken as $E=6.6\times 10^4\ \rm  V/cm$, which corresponds to a peak laser intensity  $I=c E^2\simeq 1.1\times 10^7\ \rm W/cm^2$. This particular set  
of parameters has been chosen in such a way that, for the massive model with $\Delta \varepsilon=0.26\ \rm eV$, the single-particle distribution function $W_\mathpzc{g}(\pmb{0},t)$  hits a resonance corresponding to the absorption of $n\approx 15$ photons from 
the strong OEF [see Eq.~(\ref{rensonantcondition})]. At this point, we remark that similar field  parameters are  comfortably attainable with terahertz laser pulses  of  picosecond 
duration \cite{tera1,tera2}. Likewise, the presence of a critical field $E_\mathpzc{g}$ establishes a typical scale of  intensity 
in graphene  $I_\mathpzc{g}=cE_\mathpzc{g}^2$ which amounts to $I_\mathpzc{g}\simeq 1.8 \times 10^8\ \rm W/cm^2$ for $\Delta \varepsilon=0.26\ \rm eV$. This intensity level  can be easily approached and overpassed with the current laser technology.

\begin{figure}
\includegraphics[width=0.52\textwidth]{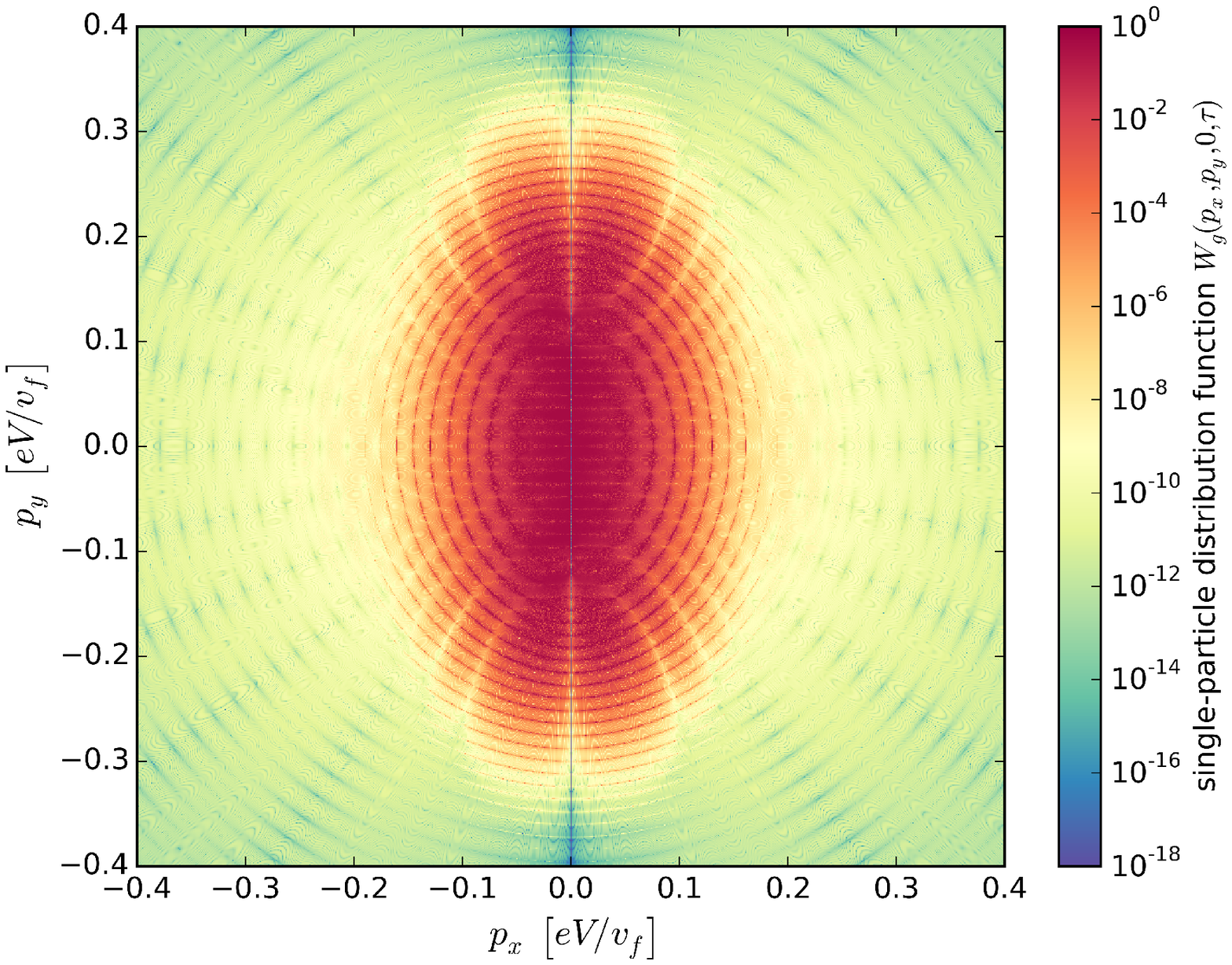}
\includegraphics[width=0.52\textwidth]{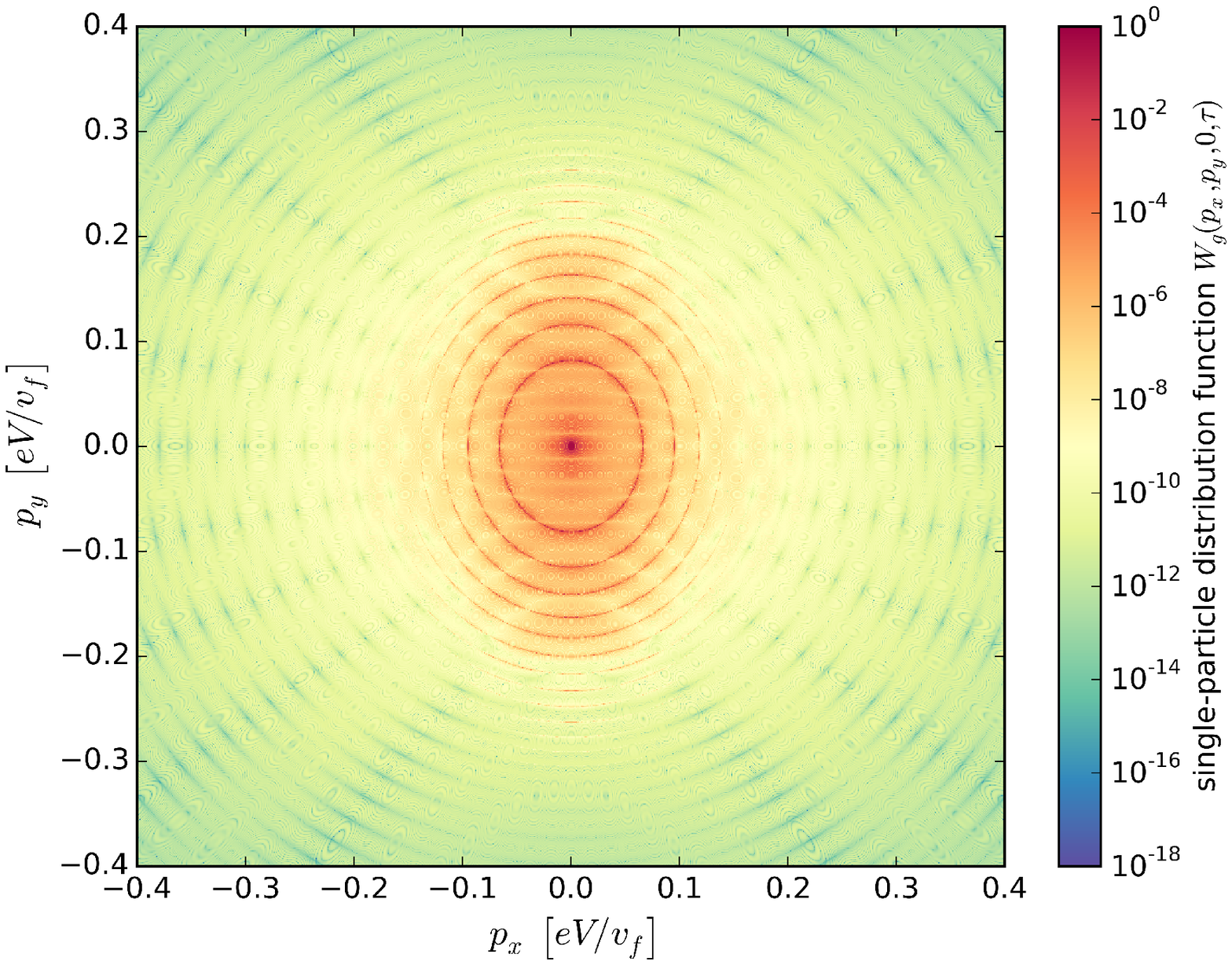}
\caption{Logarithmic plots of  the  single-particle distribution functions for models involving massless  [upper panel] and massive [$\Delta \varepsilon=0.26\ \rm eV$, lower panel] charged carriers are displayed. 
In both cases the external parameters were chosen as follows:  $E=6.6\times 10^4\ \rm V/cm$, $\tau \simeq 41.625\ \rm ps$ and $\omega=24.032\ \rm meV$.}
\label{fig:1}
\end{figure}

The system of differential equations [Eqs.~(\ref{firstequa}) and (\ref{secondequa})] has been solved  varying  the  momentum components  in the range
$- 0.4\  \mathrm{eV}\leqslant p_{x,y}\mathpzc{v}_f\leqslant 0.4\  \mathrm{eV}$. The results of this assessment  are  displayed in Fig.~\ref{fig:1}  
in a density color scheme which corresponds to $\log_{10}[W_\mathpzc{g}(\pmb{p},\tau)]$.  While the upper panel shows for comparison the  outcomes  associated with 
a  massless model [$\mathpzc{m}=0$], the effects coming from the chiral symmetry breaking  are displayed  in the lower one. Manifestly, the spectral density associated 
with massive particles  looks different  as compared with the model of  massless carriers. For instance, at  zero momenta [$\pmb{p}=0$], the 
distribution function associated with massive carriers  reaches the maximum value [$W_{\mathpzc{g}}(0,\tau)=1$], whereas  the corresponding result for 
the  massless model is  minimum   [$W_{\mathpzc{g}}(0,\tau)=0$]. This minimal  value extends along the vertical line located at $p_x=0$,  which constitutes an inherent feature 
of this scenario.  Its occurrence can be  anticipated  already  in Eq.~(\ref{vlasovgraphene}) by noting that its  right-hand side is proportional to 
the transverse energy squared [$\epsilon_\perp^2=p_x^2\mathpzc{v}_f^2+\mathpzc{m}^2\mathpzc{v}_f^4$], which vanishes identically when the  carriers are massless 
and  $p_x\to0$. In such a situation, the kinetic equation reduces to $\dot{W}_\mathpzc{g}(p_y,t)=0$ and the only conceivable solution--in accordance with 
our  initial condition--is $W_\mathpzc{g}(p_y,t)=0$. 

The ringlike structures displayed in each panel  are  understood as  isocontours of quasienergy $\bar{\varepsilon}_{\pmb{p}}$ satifying the  resonance  condition 
[Eq.~(\ref{rensonantcondition})] and, accordingly, the  number of photons involved in each resonant process can be inferred.  So, further insights on  $W_\mathpzc{g}(\pmb{p},\tau)$  
can be acquired by contrasting  the  numerical results  with the approached behavior near a resonance  [Eq.~(\ref{resonantdistributionfunction})]. To this end,  $W_{\mathpzc{g},n}(\pmb{p},t)$ 
is  evaluated  for times larger  than the interaction time [$t>\tau$], when it has become constant.  At the resonance, i.e., where  
Eq.~(\ref{rensonantcondition}) holds  [$\Delta_{n}\simeq0$], the Rabi-like frequency  reduces to $\Omega_{\mathrm{Rabi}}(\pmb{p})\approx \frac{1}{2} \vert\Lambda_{n}(\pmb{p})\vert$ and 
\begin{eqnarray}\label{rabioscillationvacuumquasiparticle}
W_{\mathpzc{g},n}(\pmb{p},\infty)\approx \sin^2\left[\Omega_{\mathrm{Rabi}}(\pmb{p})\tau\right].
\end{eqnarray} For a given interacting time  $\tau$,  Eq.~(\ref{rabioscillationvacuumquasiparticle})  achieves its maximum value [$W_{\mathpzc{g},n}(\pmb{p}_0,t)=1$]  
for a certain momenta combination $\pmb{p}_0=(p_{x0},p_{y0})$  provided $\Omega_{\mathrm{Rabi}}(\pmb{p}_0)=(2k+1)\pi/\tau$ with $k\in\mathbb{Z}$ is satisfied.  Clearly, 
our previous discussion indicates that the former has been chosen  so that the maximum is achieved at $\pmb{p}=0$.  Observe that, away from the resonance, i.e. for 
$\Delta_{n}\neq0$,  the  amplitude of $W_{\mathpzc{g},n}(\pmb{p},\tau)$ decreases [see Eq.~(\ref{resonantdistributionfunction})]. This trend  manifests in both panels 
around each resonance in  form of  light red--sometimes orange--valleys.  Additionally, they reveal a gradual decrement in the single-particle distribution functions as the
momentum components increase.

\begin{figure}
\includegraphics[width=.48\textwidth]{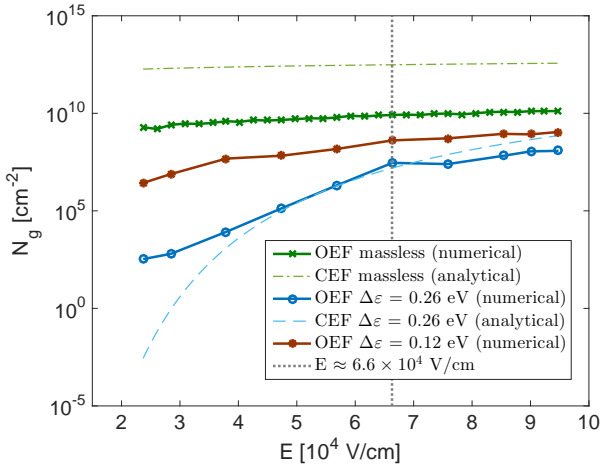}
\caption{Number of electron-hole  pairs produced per $\rm cm^2$ in an OEF.  While the result for massless carriers is in green, the outcome for massive charges is shown in red and blue, with the data points connected by straight lines.    
Also, for comparison, the  curves  corresponding to the cases driven by a CEF [see Eq.~(\ref{pairsingraphene})] are displayed in green dash-dotted line  [for massless] and blue dashed line [for massive]. Here the vertical grey dashed 
line indicates the electric field value which was used in  Fig.~\ref{fig:1}. 
The same  benchmark values and notation of Fig.~\ref{fig:1} must be understood.}
\label{fig:2}
\end{figure}

We note that the isocontours resemble  ellipses with  major axis lying  along the longitudinal direction. The elongation along  the $p_y$-axis is an outcome of the asymmetry  introduced by 
the external field in Eq.~(\ref{vlasovgraphene}) through the term $p_y-e\mathpzc{A}(t)/c$, and indicates that the creation of a particle (resp. hole) with rather large longitudinal momentum is 
more likely to take place than with a correspondigly large transverse momentum. Such a trend resembles the pattern occurring  in a CEF, where the  distribution function is homogeneous along the longitudinal 
direction but suppressed by a Gaussian profile transversally. Furthermore, a comparison between  both panels reveals that the red-colored region in the massless model  is considerably larger than the 
corresponding region for  the massive scenario, where only a few resonances arise. This indicates that the volume below the surface $W_{\mathpzc{g}}(p_x,p_y)$ for  the former  exceeds the one 
associated with the cases driven by massive carriers. Since both  volumes result from the  integration  over the momentum components,  actually  they  will  determine the number of pairs 
yielded  for each scenario  in a vicinity of a  Dirac point. Hence, the results shown in Fig.~\ref{fig:1}  already provide evidence that the density of created pairs in the massless 
case is substantially larger than the one resulting from the situation  dealing with massive carriers.  

The previous claim is verified in Fig.~\ref{fig:2}, where the behavior of  Eq.~(\ref{Ngraphene}) with respect to the electric field strength $E$ is shown.  In this figure, the  green curve 
describes the number of pairs yielded per $\rm cm^2$ for  massless charges, whereas the one in blue corresponds to the  massive carrier model with $\Delta\varepsilon=0.26 \ \rm eV$. 
For comparison, the respective outcomes in a CEF have been included. Figure~\ref{fig:2}  shows that the production efficiency for the case of massive particles is strongly reduced by several 
orders of magnitude. This mainly reflects the impact of the Schwinger-like exponential factor in Eq.~(\ref{pairsingraphene}) which is absent when the particles are massless. Accordingly, the 
curves for the massless case are rather flat, whereas the curves for massive Dirac fermions exhibit a much stronger dependence on the applied electric field strength.

For the case of massive particles with $\Delta \varepsilon=0.26\ \rm eV$, our numerical results for an OEF and the analytical prediction for a CEF lie quite close to each other at field strengths above 
$E\gtrsim  4\times 10^4$ V/cm. This may be understood by observing that the field oscillations are rather slow [$\omega \ll \mathpzc{m}$] such that the OEF locally resembles a CEF. Besides, the dimensionless 
intensity parameter $\xi_{\mathpzc{g}}=\vert e\vert E/(\mathpzc{m}\omega \mathpzc{v}_f)$  is of order unity here [$\xi_{\mathpzc{g}}=1$ corresponds to $E \approx 4.7\times 10^4$ V/cm for the gap value chosen $\Delta\varepsilon=0.26\ \rm eV$]. 
Contrary to that, for smaller field strengths below $E\lesssim 3\times 10^4$ V/cm, the pair  density produced by an OEF is significantly larger than the CEF outcome. The reason is that in an OEF pairs can be 
generated both by the field amplitude but also by the time dependence of the field \cite{Avetissian,Fillion-Gourdeau:2015dga}; the latter channel is absent in a CEF. Accordingly, for $\xi_{\mathpzc{g}}<1$, the 
creation mechanism by multiphoton absorption can become dominant, leading to enhanced pair creation in an OEF as compared with a CEF.

For the case of massless particles, our numerical results for an OEF and the analytical prediction for a CEF in  Fig.~\ref{fig:2}  run almost parallel. However, the CEF outcome--which describes the saturation density in accordance 
with Eq.~(\ref{pairsingraphene})--is larger by a few orders of magnitude. We argue that this is because the field oscillations can be considered as very fast in this case  [$\omega \gg \mathpzc{m}=0$]. Hence, the effective field strength 
acting during the pair formation time is reduced by a corresponding temporal average, leading to the strong suppression observed in  Fig.~\ref{fig:2}. We point out that this behavior is different in a non-oscillating, Sauter-like 
electric field whose pair creation efficiency of massless charge carriers in graphene approaches the CEF result to within a factor of order unity \cite{Mostepanenko}.

For comparison, Fig.~\ref{fig:2} also shows our estimates for the  number of  massive pairs produced in graphene with a reduced gap of  $\Delta\varepsilon=0.12\ \mathrm{eV}$. As one might have expected, the number is significantly 
larger than in case of $\Delta\varepsilon=0.26\ \mathrm{eV}$ and approaches to the massless limit.  We note that the critical  field in the case presently under consideration only amounts to $E_\mathpzc{g}\simeq 5.5\times 10^4\ \rm V/cm$. Since this turns 
out to be  comparable with the values encompassed in the picture, the analytical expression associated with a constant electric field does not apply and the corresponding curve  is not shown here.

\section{Conclusion and outlook}

In summary, we have started by presenting the fundamental aspects associated with the production of pairs in a $3+1$-dimensional spacetime 
for the specific case  in which the external gauge potential  is independent of the space coordinates but oscillates in time.  With the aim 
of investigating PP  in low dimensional Minkowski spaces,  dimensional  compactifications  have  been carried out. The resulting effective 
theories  verify that the production process  in  $2+1$- and $1+1$-dimensional spacetime is  appropriately described by  quantum-kinetic 
Boltzmann-Vlasov equations. 

We have applied the former outcome  to the production of massive Dirac fermions in graphene layers. The  situation 
driven by an electric field oscillating  in time has been investigated, which is characterized by a pronounced resonant behavior. The latter 
is reflected in the momentum distribution of created pairs. The total number density of massive pairs shows a strong dependence on the applied 
electric field strength which is mainly due to the Schwinger-like exponential factor involving the critical field amplitude in graphene. Its 
occurence is an important consequence of the massive quasiparticles in bandgap graphene layers, as compared with the massless charge carriers 
in ordinary gap-free graphene. Differences between the pair densities obtained in oscillating and constant electric fields, respectively, 
could be traced back to the impact of the field oscillations.

Our numerical findings have shown that terahertz laser pulses--in  combination with the  substrate-induced bandgap technique for graphene--might 
provide a feasible scenario in which the creation of light quasiparticle-hole pairs  could take place;  this way simulating the  vacuum instability in 
QED.  In connection, it remains an open question whether the 
production rate of massless quasiparticles in a single mode field can be comparable to the results associated with light massive Dirac fermions when 
a weak beam of energetic photons is superimposed onto a strong but low-frequency electric field.  We plan to present a detailed investigation of 
this  subject  in a forthcoming publication.

\section*{Acknowledgments}
We thank Thomas Heinzel, Alexander B. Voitkiv, and Hengyi Xu for useful discussions. 
S. Villalba-Ch\'avez  gratefully acknowledges the support of  the Alexander
von Humboldt Foundation in the early stage of this project.  R. Egger  acknowledges funding by the network SPP 1459 of the German Research Foundation (DFG).

\end{document}